\documentclass[usenatbib]{mn2e}
\usepackage[utf8]{inputenc}
\usepackage{natbib,epsfig,amsmath,cite,amssymb,graphicx,mystyle}
\usepackage{mathtools}
\usepackage{array, xcolor, caption, multirow}
\usepackage[colorlinks=true,linkcolor=blue,citecolor=blue]{hyperref}

\title[Stellar evolution and GW observations]{Exploring stellar evolution with gravitational-wave observations}

\author[Dvorkin, Uzan, Vangioni, Silk]{Irina Dvorkin$^{1,2}$\thanks{E-mail: irina.dvorkin@aei.mpg.de}, Jean-Philippe Uzan$^{2}$, Elisabeth Vangioni$^{2}$, Joseph Silk$^{2,3,4}$\\
$^{1}$ Max Planck Institute for Gravitational Physics (Albert Einstein Institute), Am M\"{u}hlenberg 1, Potsdam-Golm, 14476, Germany\\
$^{2}$ Institut d'Astrophysique de Paris, Sorbonne Universit\'{e}s, UPMC Univ Paris 6 \& CNRS, UMR 7095, 98 bis bd Arago,\\ F-75014 Paris, France\\
$^{3}$ Department of Physics and Astronomy, The Johns Hopkins University,
Baltimore, MD 21218, USA \\
$^{4}$ BIPAC, University of Oxford, 1 Keble Road, Oxford OX1 3RH, UK \\
}

\begin{document}
\newcommand{\dd}{{\rm d}}

\pagerange{\pageref{firstpage}--\pageref{lastpage}} \pubyear{2017}
\maketitle
\label{firstpage}

\begin{abstract}
Recent detections of gravitational waves from merging binary black holes opened new possibilities to study the evolution of massive stars and black hole formation. In particular, stellar evolution models may be constrained on the basis of the differences in the predicted distribution of black hole masses and redshifts. In this work we propose a framework that combines galaxy and stellar evolution models and use it to predict the detection rates of merging binary black holes for various stellar evolution models. We discuss the prospects of constraining the shape of the time delay distribution of merging binaries using just the observed distribution of chirp masses. Finally, we consider a generic model of primordial black hole formation and discuss the possibility of distinguishing it from stellar-origin black holes. 
\end{abstract}


\section{Introduction}
The discovery of the first gravitational-wave (GW) source GW150914, a merger of two black holes (BHs), by Advanced LIGO \citep{2016PhRvL.116f1102A} marked the birth of a new astronomical discipline. Analysis performed on the first two Advanced LIGO observing runs (the last part of the second run being conducted in parallel with Advanced Virgo) has so far resulted in the detection of four additional sources, as well as a tentative, lower-significance, candidate event \citep[GW151226, LVT151012, GW170104, GW170814; GW170608; ][]{2016PhRvX...6d1015A,2017PhRvL.118v1101A,2017arXiv170909660T,2017arXiv171105578T}. These observations have notably shown for the first time that heavy ($\gtrsim 20M_{\odot}$) BHs exist and can form binaries that merge within the age of the Universe. Furthermore, the joint observation of GW170814 by Advanced LIGO and Advanced Virgo demonstrated the added accuracy (a reduction of over an order of magnitude in positional uncertainty) that can be reached with three detectors \citep{2017arXiv170909660T}.
As the sensitivity of ground-based interferometers increases, future GW observations of merging BH binaries will provide more precise information on their masses, spins and redshifts. Indeed, it is expected that a few tens to a few hundreds of events will be observed within the next several years \citep{2016PhRvX...6d1015A}. This wealth of data can of course be used to study the models that describe how BHs form. 

The leading scenario that has been proposed to explain the formation of stellar-mass ($\lesssim 100M_{\odot}$) BHs relies on the standard evolution channel of massive ($\gtrsim 20M_{\odot}$) field stars. After the iron core collapses, a BH can form either after a supernova explosion and the following (partial) fallback or matter and eventual collapse, or a direct collapse of the entire stellar envelope \citep{1995ApJS..101..181W,2010ApJ...715L.138B,2012ApJ...749...91F,2017arXiv170601913L}. An interesting phenomenon occurs in the mass range of $\sim 130-250M_{\odot}$ (but note the dependence on metallicity and rotation velocity) where the star becomes unstable due to production of electron-positron pairs and undergoes a pair-instability supernova (PISN). In this case the star is completely disrupted and no remnant is left \citep{1964ApJS....9..201F}. While the conditions that lead to, or prevent, a successful supernova explosion are not yet fully understood \citep[see e.g.][]{2011ApJ...730...70O,2012ApJ...761...72M}, the evolution of \emph{binary} massive stars is even less certain. The binary orbit is thought to decay during a common envelope phase \citep{2001ASPC..229..239P,2013A&ARv..21...59I} with a possible contribution from a chemically homogeneous evolution channel \citep{2016MNRAS.458.2634M,2016A&A...588A..50M}. A complementary channel for binary BH formation, driven by mergers in dense stellar environments, may become dominant in stellar clusters \citep[e.g.][]{2014MNRAS.441.3703Z,2016ApJ...831..187A,2016PhRvD..93h4029R,2017PhRvD..95l4046G,2017arXiv170902058F}. Other possible scenarios for forming stellar-mass binary BHs include primordial BHs \citep{2016PhRvL.116t1301B,2016PhRvL.117f1101S} and population III remnants \citep{2014MNRAS.442.2963K,2016MNRAS.460L..74H,2016MNRAS.461.2722I}. The distributions of masses, spins and redshifts of detectable sources in each of these channels are different, which opens the possibility of studying them with upcoming GW observations. However, the number densities of sources also depend on the underlying galaxy evolution model, for example the star formation rate (SFR), which renders model selection rather challenging.

Nevertheless, several groups have recently started to explore the full potential of GW observations for stellar evolution modeling, in particular for constraining the parameters of specific models \citep[e.g.][]{2016A&A...594A..97B,2017arXiv170901943W,2017arXiv171106287B,2017MNRAS.472.2422M} as well as model selection \citep{2017arXiv170407379Z,2017PhRvD..95l4046G,2017arXiv170708978H} and direct probing of the BH mass function \citep{2017PhRvD..95j3010K}. Notably, the important issue of the properties of galaxies that host binary BH mergers has been discussed by \citet{2016MNRAS.463L..31L,2017arXiv170506781S} and \citet{2017arXiv171109190C}.

In this article we propose a general framework for the analysis of future GW observations. Our ultimate goal is to be able to constrain a large variety of stellar evolution scenarios which will be embedded in our galaxy evolution model. For the latter we use the model developed in \citet{2016MNRAS.461.3877D} and \citet{2016PhRvD..94j3011D} \citep[based on][]{2004ApJ...617..693D,2006ApJ...647..773D,2015MNRAS.447.2575V} and implement several stellar evolution models that we wish to compare. We then estimate the number of detections that would be made by LIGO in each case, as well as the mass and redshift distribution of these \emph{detectable} mergers. To demonstrate the utility of this approach we estimate the precision with which some of the model parameters can be measured with mock observations that we draw from out binary black hole populations.
Our semi-analytic approach differs from previous studies in that it will allow us to marginalize over many astrophysical 'nuisance parameters', such as the star formation rate (in particular at high redshifts, where it is poorly constrained), the time to coalescence of binary black holes etc. In other words, we can in principle treat different stellar evolution models within the same galaxy evolution scenario while simultaneously varying also the galaxy evolution parameters.

The structure of this paper is as follows. Section \ref{sec:detrates} describes our calculation of detection rates of binary BH mergers. Section \ref{sec:astro} details our galaxy evolution model as well as the four stellar evolution models which we implement here and a generic primordial black hole formation scenario. Our results for the mass and redshift distribution of \emph{detectable} mergers are presented in Section \ref{sec:res}. We then use our framework to predict the accuracy with which some of the parameters can be measured with future detections in Section \ref{sec:stats}. Finally, we discuss future applications of our framework in Section \ref{sec:dis}.

\section{Detection rates}
\label{sec:detrates}

We start with a model (to be specified below) that provides the total birth rate of BHs per unit \emph{observer} time per unit comoving volume $V$ and per unit BH mass $m$:
\begin{equation}
 \frac{\dd \dot{n}_{\rm tot}}{\dd m} = \frac{\dd N}{{\dd t_{\rm obs}}{\dd V}{\dd m}}\:.
\end{equation}
We then assume that only a fraction $\beta(m)$ of these BHs reside in binary systems that coalesce within a Hubble time:
\begin{equation}
 \frac{\dd \dot{n}_{2}}{\dd m}(m)=\beta(m) \frac{\dd \dot{n}_{tot}}{\dd m}\:.
 \label{eq:beta}
\end{equation}
Then the birth rate of binaries with component masses $m$ and $m' \leq m$ reads:
\begin{equation}
 \frac{\dd^2 \dot{n}_{\rm bin}}{{\dd m}{\dd m'}}(m,m') = \frac{\dd \dot{n}_{2}}{\dd m}\frac{\dd \dot{n}_{2}}{\dd m'}P(m',m)
 \label{eq:dndmdm}
\end{equation}
where the function $P(m',m)$ is normalized so that:
\begin{equation}
 \int \frac{\dd \dot{n}_{2}}{\dd m}\frac{\dd \dot{n}_{2}}{\dd m'}P(m',m) \dd m'\dd m = \frac{1}{2}\int \frac{\dd \dot{n}_{2}}{\dd m}\dd m \: .
\end{equation}
 
If the binary merges within a time $t_{\rm delay}$ after it has formed, where the latter is given by the normalized probability distribution $P_{\rm d}(t_{\rm delay})$: 
\begin{equation}
 \int_{t_{\rm min}}^{t_{\rm max}}P_{\rm d}(t_{\rm delay}){\dd t_{\rm delay}} =  1 \:,
\end{equation}
then the number of binaries merging per unit time $t_{\rm merge}=t+t_{\rm delay}$ is given by:
\begin{equation}
 \frac{\dd N}{{\dd t_{\rm merge}} {\dd m} {\dd m'}}=\int \frac{\dd^2 \dot{n}_{\rm bin}(t)}{{\dd m}{\dd m'}}P_{\rm d}(t_{\rm merge}-t)\frac{\dd V}{\dd z}{\dd z}{\dd t_{\rm obs}}\:.
 \label{eq:dNdt_m}
\end{equation}
In the last expression, the birth time $t$ and the corresponding redshift $z$ are related by
\begin{equation}
 \left|\frac{\dd t}{\dd z}\right|=\frac{1}{H_0\sqrt{\Omega_m(1+z)^3+\Omega_{\Lambda}}(1+z)}
 \label{eq:dtdz}
\end{equation}
and $t_{\rm obs}$ is the observation time. Since the total observation time is very short compared to cosmological scales ($T_{\rm obs}\sim 50$ days for LIGO O1), the integral over $\dd t_{\rm obs}$ is trivial. In order to obtain the number of events detectable by a given instrument, e.g. Advanced LIGO, we need to calculate the signal-to-noise rate (SNR) for each of these events:
\begin{equation}
 \rho^2=4\int \frac{|h(f)|^2}{S_n(f)}\dd f
 \label{eq:SNR}
\end{equation}
where $h(f)$ is the GW strain in the observed frequency domain and $S_n(f)$ is the noise power spectral density. Note that the strain is a function of the binary parameters: component masses and spins, redshift, orientation and sky localization. We obtain the number of observed events (defined here as those with $\rho>8$) by first calculating $P(\rho>8|m_1,m_2,z)$, the probability that a merger of BHs with masses $m_1$ and $m_2$ at redshift $z$ is detectable. We average over source orientation and component spins (assuming spins uniform in magnitude and isotropic in direction). It follows that the number of sources detectable after observing for a total time $T_{\rm obs}$ is:
\begin{multline}
 \frac{\dd N_{\rm det}}{{\dd t_{\rm merge}} {\dd m} {\dd m'}}=T_{\rm obs}\int \frac{\dd^2 \dot{n}_{\rm bin}}{{\dd m}{\dd m'}}P_{\rm d}(t_{\rm merge}-t)\\P(\rho>8|m,m',z_{\rm merge})\frac{\dd V}{\dd z}{\dd z}\:.
 \label{eq:Nobs}
\end{multline}

In this work we assume the following distributions:
\begin{equation}
 P(m',m) = \textrm{constant} ,\: m,m'\in [M_{\rm min},M_{\rm max}]
 \label{eq:Pmm}
\end{equation}
and
\begin{equation}
 P_{\rm d}(t_{\rm delay})\propto t^{-\gamma}_{\rm delay},\: t\in [t_{\rm min},t_{\rm max}]
 \label{eq:tdist}
\end{equation}
with $t_{\rm min}=50$ Myr and $t_{\rm max}=t_H$, where $t_H$ is the Hubble time \citep{2012ApJ...759...52D}. The specific form of the function $P(m',m)$ was adopted here for simplicity, other choices will be explored in future work. Furthermore we assume that the fraction of BHs that are in binaries and that merge within a Hubble time is $\beta$ and does not depend on mass. We take $\gamma=1$ \citep[e.g.][]{2013ApJ...779...72D} and $\beta=0.01$ as fiducial values  and explore the possibilities of constraining them with LIGO observations in Section \ref{sec:stats}.

In order to calculate the SNR from eq. (\ref{eq:SNR}) we use the \emph{PhenomB} inspiral-merger-ringdown waveforms \citep{2011PhRvL.106x1101A} and the noise power spectral density from \citet{2016PhRvL.116m1103A}. 

In order to compare our model predictions to observational data we present below the detection rate in the primary mass-secondary mass plane, in units of $M_{\odot}^{-2}yr^{-1}$:
\begin{equation}
 R_{\rm det}(m,m') = \frac{1}{T_{\rm obs}}\int \frac{\dd N_{\rm det}}{{\dd t_{\rm merge}} {\dd m} {\dd m'}}{\dd t_{\rm merge}}\:.
 \label{eq:Rdetdef}
\end{equation}

In the next Section we discuss the astrophysical models that provide the birth rate of binary BHs.

\section{Astrophysical models}
\label{sec:astro}

\subsection{Galaxy evolution}

There are two astrophysical terms in eq. (\ref{eq:Nobs}): the birth rate of binaries ${\dd \dot{n}}/{\dd m} {\dd m'}$ and the probability to merge after a time delay $t_{\rm delay}$ given by $P(t_{\rm delay})$. Some of the current stellar evolution models can predict the birth rate of binaries with a certain set of orbital parameters, from which the merging time due to emission of GW can be calculated \citep{2010ApJ...715L.138B,2016Natur.534..512B}. 
Other models provide only the birth rate ${\dd \dot{n}}/{\dd m} {\dd m'}$ and have to rely on some distribution of merging times $P(t_{\rm merge})$. Moreover, most astrophysical models utilize some distribution of the component masses of the stellar binary as an input. It should also be kept in mind that the birth rate of BHs follows from the formation rate of their progenitor massive stars and so depends on the global star formation rate and the stellar initial mass function, as well as stellar metallicity and local density (for example, multiple mergers can occur in dense stellar environments). Therefore, the stellar evolution model that we wish to test needs to be embedded in a galaxy evolution framework, either (semi-)analytical or numerical.

In this work we rely on the semi-analytic approach developed in \citet{2016MNRAS.461.3877D} and \citet{2016PhRvD..94j3011D}, which is based on the galaxy evolution model in \citet{2004ApJ...617..693D,2006ApJ...647..773D} and \citet{2015MNRAS.447.2575V}. To sum up, our model takes as an input the structure formation history (computed with the Press-Schechter semi-analytic approach), the star formation rate (SFR) history, the initial mass function and stellar yields. Another crucial input is the relation between initial stellar mass and metallicity and the remnant (neutron star or black hole) mass. The latter component is taken from detailed stellar evolution models that we want to test, as described below. The output of our model is the evolution of the chemical composition of the interstellar and circumgalactic media and the number densities of black holes and neutron stars, as well as other astrophysical quantities, i.e. gas fraction and the optical depth to reionization, used to calibrate the model. We assume the Salpeter stellar initial mass function \citep{1955ApJ...121..161S} in the mass range $0.1-100M_{\odot}$ and calibrate our SFR to the observations compiled by \citet{2013ApJ...770...57B}, complemented by those by \citet{2015ApJ...803...34B} and \citet{2015ApJ...808..104O}, as described in \citet{2015MNRAS.447.2575V}. We use the metal yields from \citet{1995ApJS..101..181W} for all of our models. Further discussion on the constraints on metallicity evolution and SFR, as well as a more detailed model description, can be found in \citet{2016MNRAS.461.3877D}.

\subsection{Stellar evolution and initial mass-remnant mass relation}

\begin{figure*}
\begin{tabular}{cc}
\epsfig{file=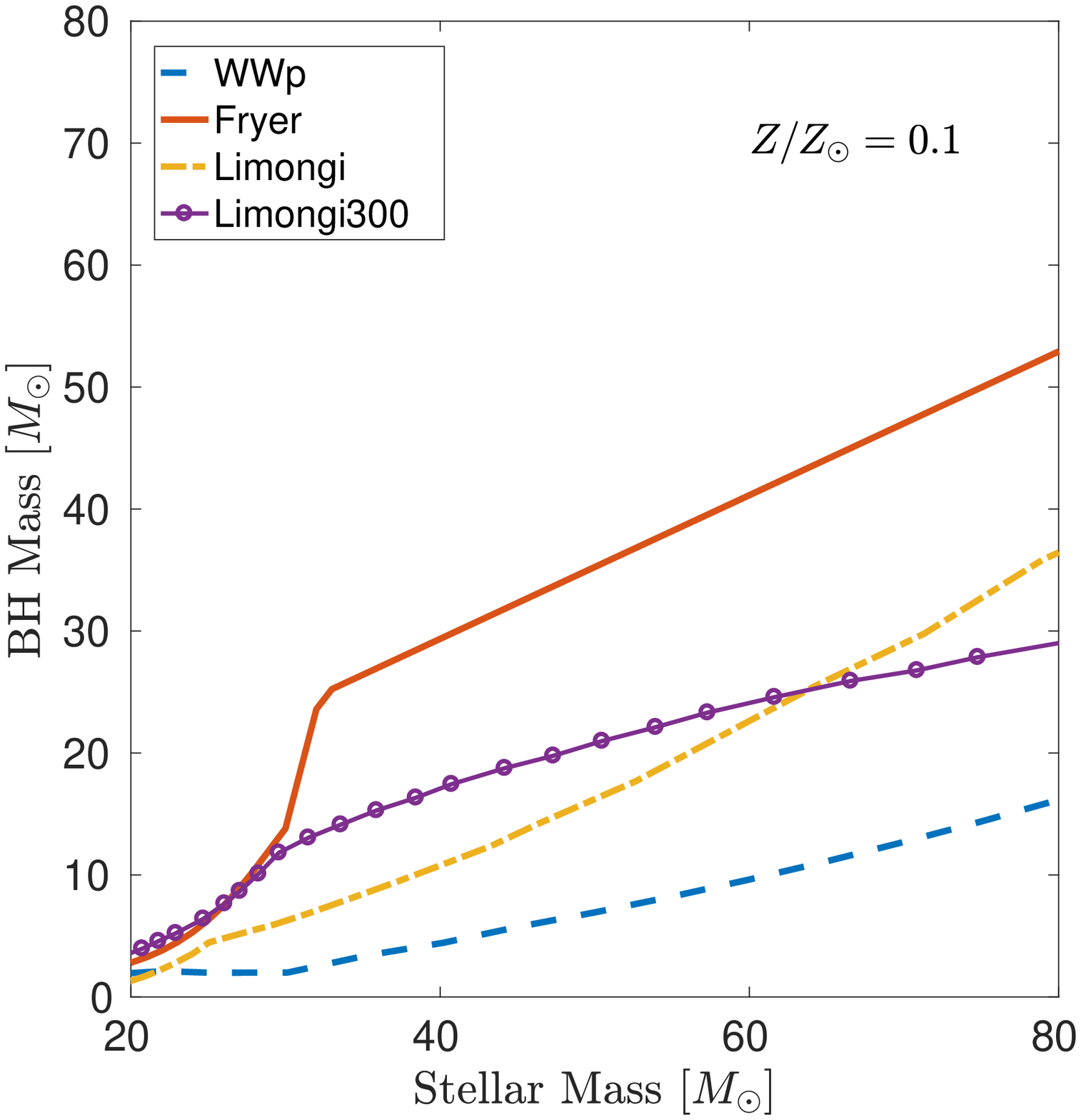, width=.4\textwidth} &
\epsfig{file=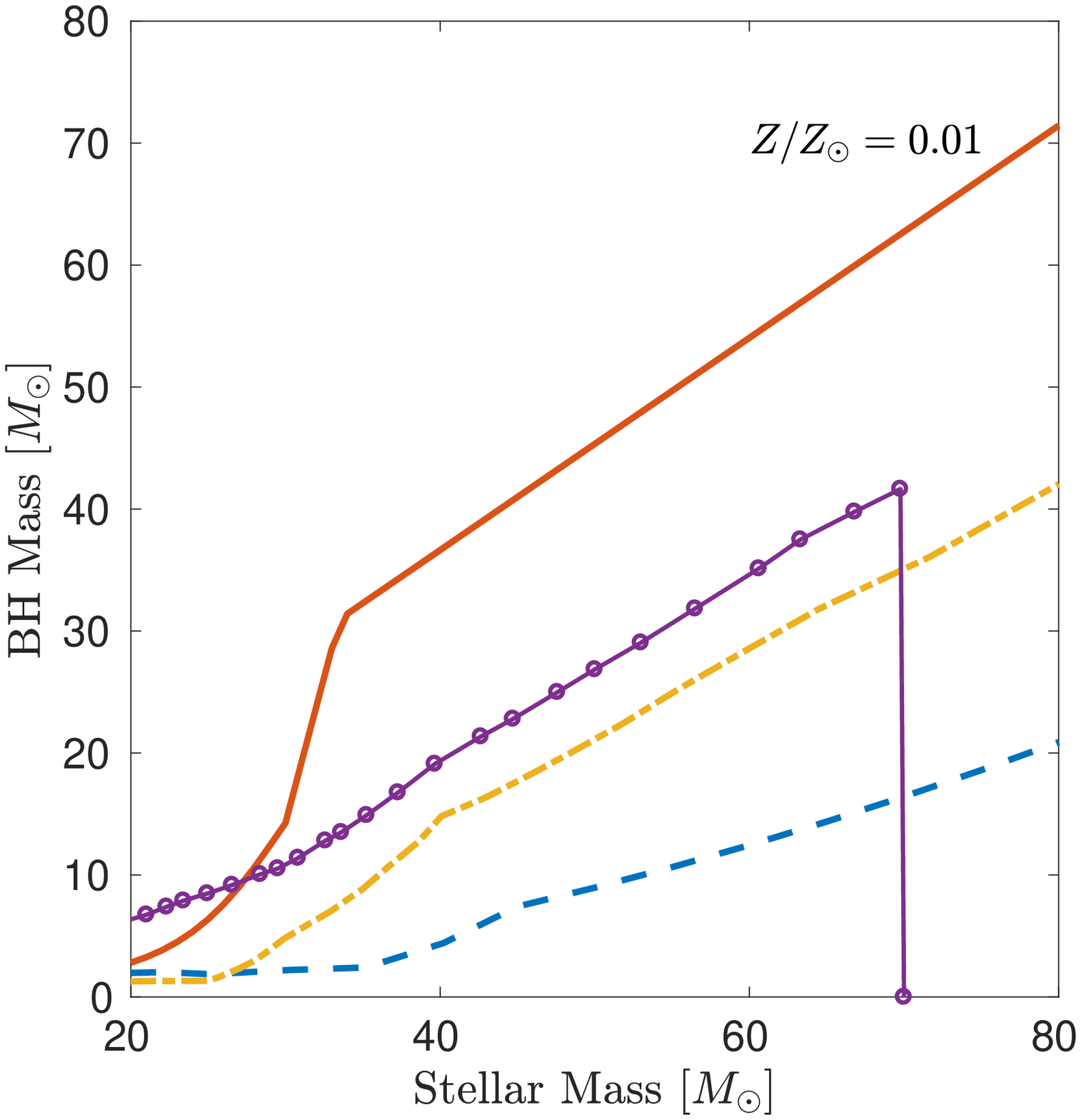, width=.4\textwidth}
\end{tabular}
\caption{Initial mass-remnant mass relation for the stellar models used in this work for two metallicity values, $Z=0.1Z_{\odot}$ (left panel) and $Z=0.01Z_{\odot}$ (right panel). Note that the BH masses are higher in the lower-metallicity case, except for the \emph{Limongi300} model which exhibits a cutoff at $M\sim 70M_{\odot}$ (see text for discussion).}
\label{fig:models}
\end{figure*}

In order to relate the initial stellar mass to the remnant mass we used four stellar evolution models: (1) the \emph{Fryer} model, based on the \emph{delayed} model in \citet{2012ApJ...749...91F}; (2) the \emph{WWp} model, based on \citet{1995ApJS..101..181W}; and (3)-(4), two models from \citet{2017arXiv170601913L} with and without stellar rotation, which we name \emph{Limongi300} and \emph{Limongi}, respectively. All of these models provide the remnant mass as a function of initial stellar mass and metallicity. Since we use  \citet{1995ApJS..101..181W} to calculate stellar yields in all of these cases, the \emph{WWp} model is the most consistent choice. Note, however, that it is based on rather old 'piston' pre-collapse stellar models and assumes a constant explosion energy. Recent studies suggested that the explosion is powered by neutrinos stored behind the shock \citep{2001ApJ...554..548F,2012ApJ...749...91F}. In this picture the explosion energy depends on neutrino heat transport mechanisms, the nature of the hydrodynamic instabilities that convert neutrino thermal energy to kinetic energy that can power the supernova \citep[e.g.][]{2003ApJ...584..971B}, and the resulting time delay between shock bounce and explosion. \citet{2012ApJ...749...91F} provide an analytic model for the latter and calculate the explosion energy, as well as the remnant mass, using numerical pre-collapse stellar models from \citet{2002RvMP...74.1015W}. Here we use the \emph{delayed} model from \citet{2012ApJ...749...91F} as a representative case. 

\citet{2017arXiv170601913L} presents a different set of models, including the cases of rotating stars. These models differ from the ones in \citet{2012ApJ...749...91F} in two aspects. First, \citet{2017arXiv170601913L} uses a different set of pre-collapse stellar models which vary from \citet{2002RvMP...74.1015W}  in their treatment of convection, mass-loss rate and angular momentum transport. For example, the metallicity dependence of the mass-loss rate used in \citet{2012ApJ...749...91F} is $\dot{M}\propto Z^{0.5}$, where $Z$ is the metallicity \citep[][]{1989A&A...219..205K}, whereas \citet{2017arXiv170601913L} use the steeper relation obtained in \citet{2001A&A...369..574V}: $\dot{M}\propto Z^{0.85}$. Second, \citet{2017arXiv170601913L} assumed a constant explosion energy in the calculation of the remnant mass, similar to the approach of \citet{1995ApJS..101..181W} and contrary to \citet{2012ApJ...749...91F}. As we will see below, these differences amount to significant discrepancies in the mass distribution of \emph{detectable} BHs among the \emph{Fryer} and \emph{Limongi} models.

Finally, the \emph{Limongi300} model allows us to test the effect of rotation of the distribution of remnant masses. Rotation affects the evolutionary path of a massive star by lowering the effective gravity and inducing rotation-driven mixing. According to the results of \citet{2017arXiv170601913L}, the main effect of rotation on the resulting BH mass is to reduce the minimal mass required for the PISN stage therefore limiting the maximal BH mass. In order to test this model we assumed that all the stars rotate at $300$ km/sec (rather than using a distribution of velocities).

The initial mass-remnant mass relation for these models is shown in Figure~\ref{fig:models}. There is a clear 'mass hierarchy' among the models, with the exception of \emph{Limongi300} which exhibits a cutoff at $M_{\rm star}\sim 70M_{\odot}$. This is the result of the fact that rotating stars enter the pair-instability regime at lower masses than non-rotating stars, as can be seen in Figs. 24g and 24i in \citet{2017arXiv170601913L}. Note also the nearly vertical relationship obtained in the \emph{Fryer} model around $M_{\rm star}\sim 30$. This is the result of the prescription for stellar winds adopted in this model (see their Eq.~(7) and Fig.~4). We will see below that this feature creates an imprint on the observed BH mass distribution. Note also that in all the cases the BH masses are higher at low metallicity, as expected in view of the reduced stellar winds.

\subsection{Primordial black holes}
\label{sec:pbh}

As first suggested by \citet{1967SvA....10..602Z,1971MNRAS.152...75H}, BHs can form during the radiation- or matter-dominated era from large primordial curvature
perturbation generated by inflation. Interestingly, this mechanism can in principle form BHs with masses ranging from the Planck mass ($10^{-5}$ g) to $\sim 10^5 M_{\odot}$, depending on their formation epoch, although BHs lighter than $\sim 10^{15}$ g would have evaporated by the present epoch \citep[see][and references within]{2018CQGra..35f3001S}. Depending on their mass, PBHs may leave observable traces that can be used to study models of the early Universe. In addition, PBHs are compelling dark matter candidates, and while a large variety of observations provide stringent constraints on the cosmological density of PBHs, certain mass ranges are still not excluded \citep{2016PhRvD..94h3504C}.

While the mechanism of PBH formation has been extensively studied in the context of inflationary models, the formation of \emph{binary} PBHs and their merger rates has received little attention until the first discovery of GW from merging $\sim 30M_{\odot}$ BHs, which raised the possibility that this was also the first detection of PBHs. Two possible mechanisms of binary formation were proposed by \citet{2016PhRvL.116t1301B} and \citet{2016PhRvL.117f1101S}, respectively. In the former scenario, PBHs constitute a significant fraction (up to $\sim 100\%$) of dark matter, and form binaries at late epochs ($z= 0$) in dense galactic environments. In the latter model, on the other hand, binaries form at early epochs via 3-body interactions. Note that these two scenarios result in very different (several orders of magnitude, depending on the PBH density) observable merger rates.

In view of the uncertainties in both these scenarios it may be useful to consider a more general phenomenological description, where PBH binaries can form (and merge) at any given epoch, which we provide in what follows.

Let us assume that all (single) PBHs were formed at the epoch of matter-radiation equality $z_{\rm eq}$ with a power-law mass function:
\begin{equation}
 \frac{\dd n}{\dd m}\propto m^{-\alpha}
\end{equation}
normalized so that they account for a fraction $q$ of the total dark matter density:
\begin{equation}
 \rho_{DM}=\frac{1}{q}\int_{M_{\rm min}}^{M_{\rm max}}\frac{\dd n}{\dd m}m{\dd m}\:.
\end{equation}
Below we will consider the case with $M_{\rm min}=10M_{\odot}$, $M_{\rm max}=1000M_{\odot}$, $q=0.01$ \citep[see][for the mass ranges of PBH that fit current observational constraints]{2017PhRvD..95d3534A,2017arXiv170906576A} and $\alpha=2$ (these values are chosen here for an illustrative purpose).

We then assume that a fraction $\Gamma_{\rm PBH}$ of these PBHs forms binaries per unit observer time:
\begin{equation}
 \frac{\dd n_2}{{\dd m}{\dd t_{\rm obs}}}\left(m,t \right)=\Gamma_{\rm PBH}(t)\frac{\dd n}{\dd m}\left(m \right)\:.
 \label{eq:Gpbh}
\end{equation}
We take the comoving number density of PBHs to be constant in time, by implicitly assuming that their merger rate is sufficiently small. Then the \emph{comoving} formation rate of binary PBHs is given by Eq. (\ref{eq:dndmdm}), where we assume $P(m',m)=\rm const.$ for $m,m'\in [M_{\rm min},M_{\rm max}]$.

To obtain the number density of \emph{mergers} we assume the following probability to merge with a delay $t_{\rm delay}$ \citep[inspired by][although note that in their case all the binaries form immediately after the formation of the BHs themselves]{2016PhRvL.117f1101S}:
 $P_{\rm d}\propto 1/t_{\rm delay}$.
Then the merger rate per unit time is given by Eq. (\ref{eq:dNdt_m}) whereas the number of detectable sources can be calculated using Eq. (\ref{eq:Nobs}).
 
Note that we still need to specify $\Gamma_{\rm PBH}$, the binary formation rate. For example, the mechanism proposed by \citet{2016PhRvL.117f1101S} corresponds to $\Gamma_{\rm PBH}\propto\delta(t_{\rm eq})$ where $\delta$ is the Dirac distribution, whereas in the scenario of \citet{2016PhRvL.116t1301B} binary formation occurs predominantly at lower redshift, after halo collapse.

\subsection{Merger rate calculation}
\label{sec:rates_calc}

In order to evaluate the total number of observed events from Eq.~(\ref{eq:Nobs}) we construct lightcones up to $z=15$ and calculate the mean expected number of events $\langle N_{bin}\rangle(t_{\rm merge},m,m')$ in bins of primary and secondary masses $m,m'$, volume shells $\frac{dV}{dz}\frac{dz}{dt}\Delta t$ (where $\Delta t=250$ Myr) and merging times $\Delta t_{\rm merge}=250$ Myr. Finally, we sum over all birth times $t_{\rm birth}$ to obtain the distribution of sources in the mass-redshift space. 

We can also calculate the \emph{observed} merger rate density from the number of actual LIGO detections and assuming a specific astrophysical model. For this purpose we use the procedure outlined in \citet{2016PhRvX...6d1015A} (see their Appendix C). Namely, if $\Lambda$ is the number of LIGO triggers of astrophysical origin, then it is related to the merger rate density through:
\begin{equation}
 \Lambda=R\langle VT_{\rm obs}\rangle
 \label{eq:Lambda}
\end{equation}
where $\langle VT_{\rm obs}\rangle$ is the population-averaged sensitive space-time volume of search (Eq. C3 in \citet{2016PhRvX...6d1015A}):
\begin{equation}
 \langle VT_{\rm obs}\rangle=T_{\rm obs}\int {\dd z}{\dd \theta}\frac{{\dd V_c}}{{\dd z}}\frac{1}{1+z}s(\theta)f(z,\theta)\:,
 \label{eq:totalRate}
\end{equation}
$s(\theta)$ is the normalized distribution function of the BH population with respect to the parameters $\theta$ (for example mass) and $f(z,\theta)$ is a selection function that gives the probability of detecting a source with parameters $\theta$ at redshift $z$ (in our case, this is the probability that a given source has SNR $\rho>8$). We stress that since $s(\theta)$ is a normalized distribution, our choice of $\beta$, the fraction of BHs that are in binaries and that merge within a Hubble time (see Eq. (\ref{eq:beta})) does not influence our results.

Note that the deduced merger rate density depends on the astrophysical model assumed for the analysis. For example, if we used an astrophysical model that predicts a negligible \emph{relative} number of $\sim 30M_{\odot}$ BHs, LIGO detections would imply a \emph{high} total merger rate to allow for the detected $\sim 30M_{\odot}$ events. Conversely, assuming a model that produces an over-abundance of $\sim 30M_{\odot}$ BHs would result in a \emph{low} overall merger rate.

\begin{figure*}
\begin{tabular}{cc}
\epsfig{file=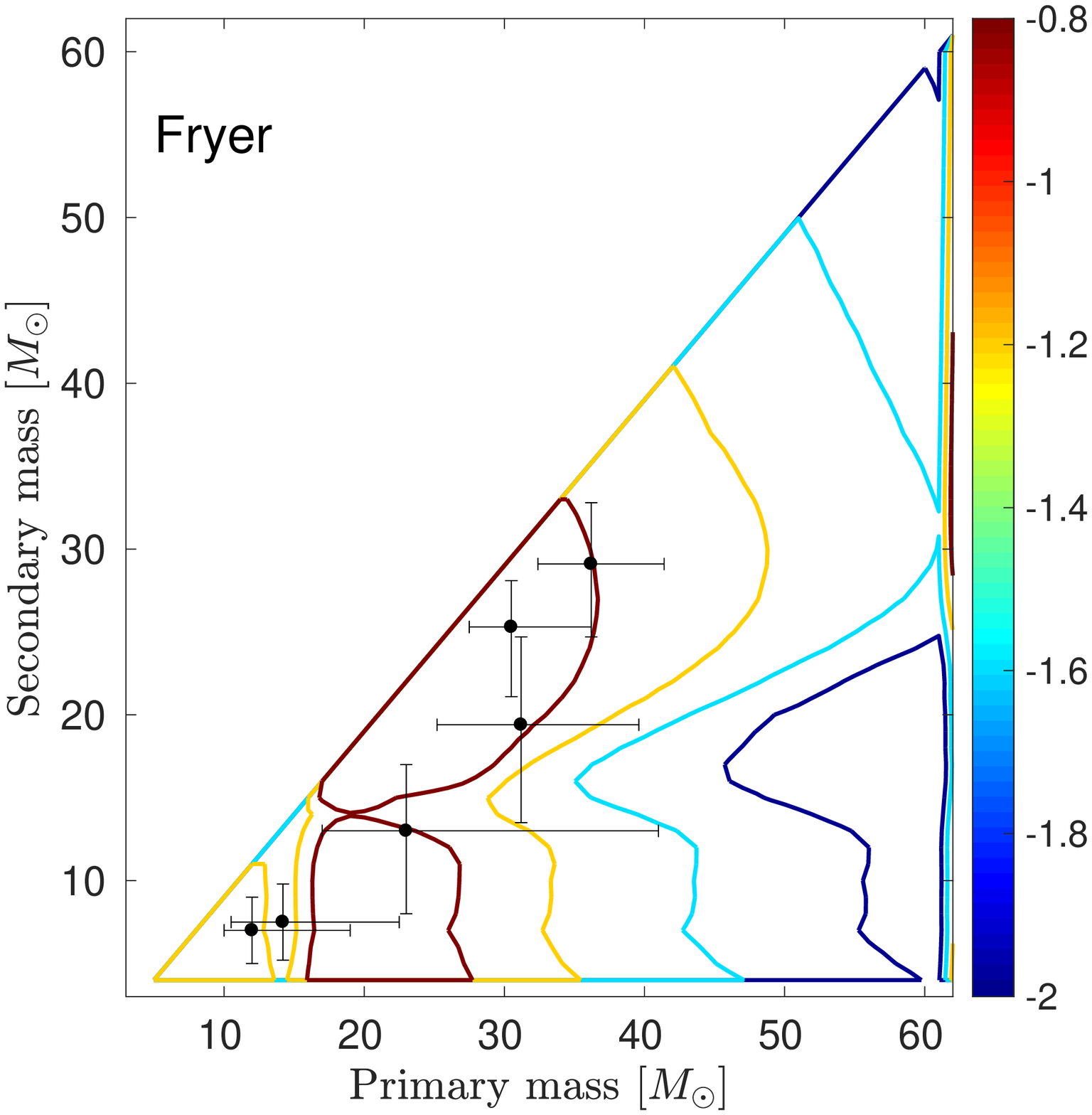, width=.3\textwidth}&
\epsfig{file=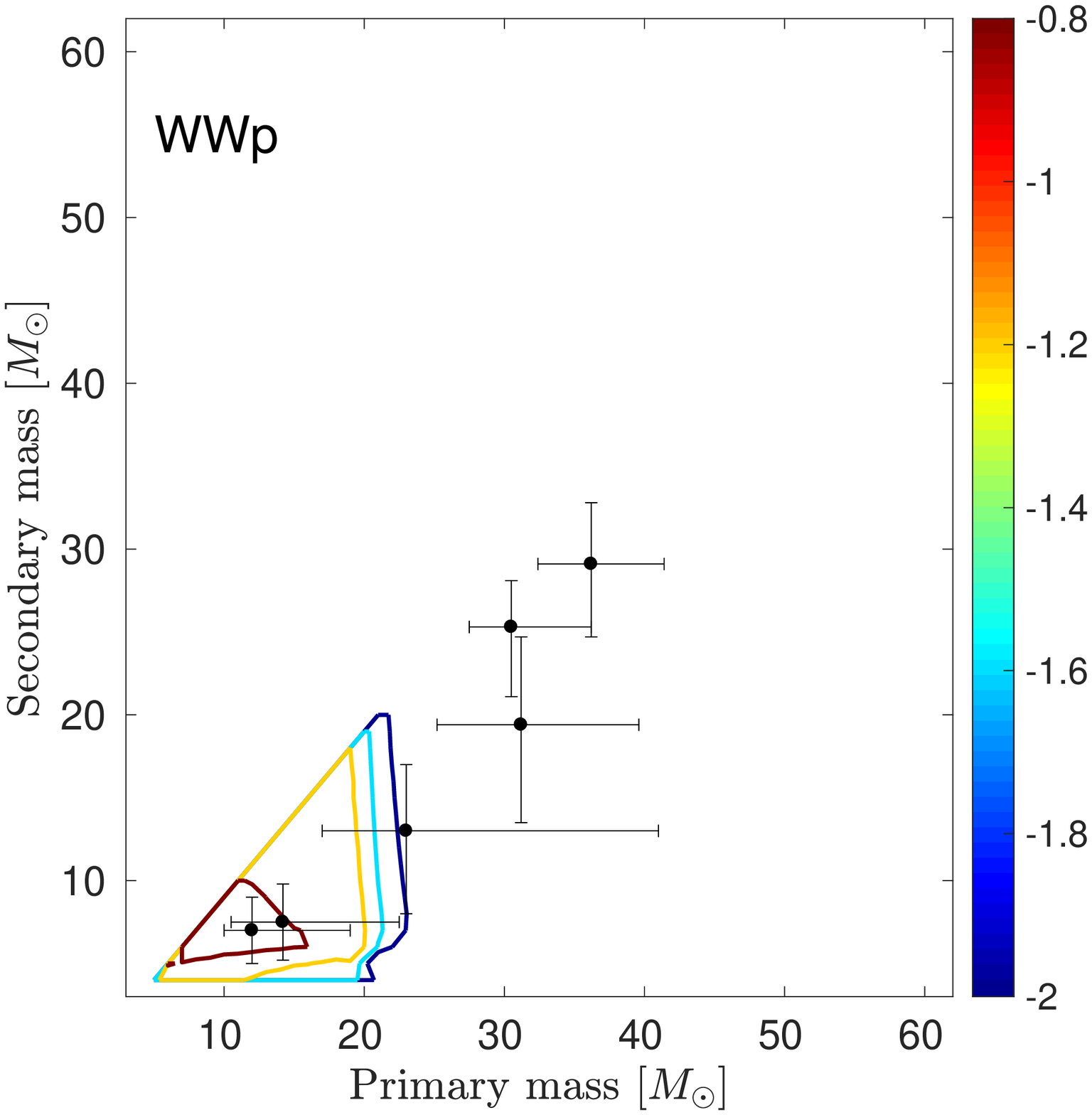, width=.3\textwidth} \\
\epsfig{file=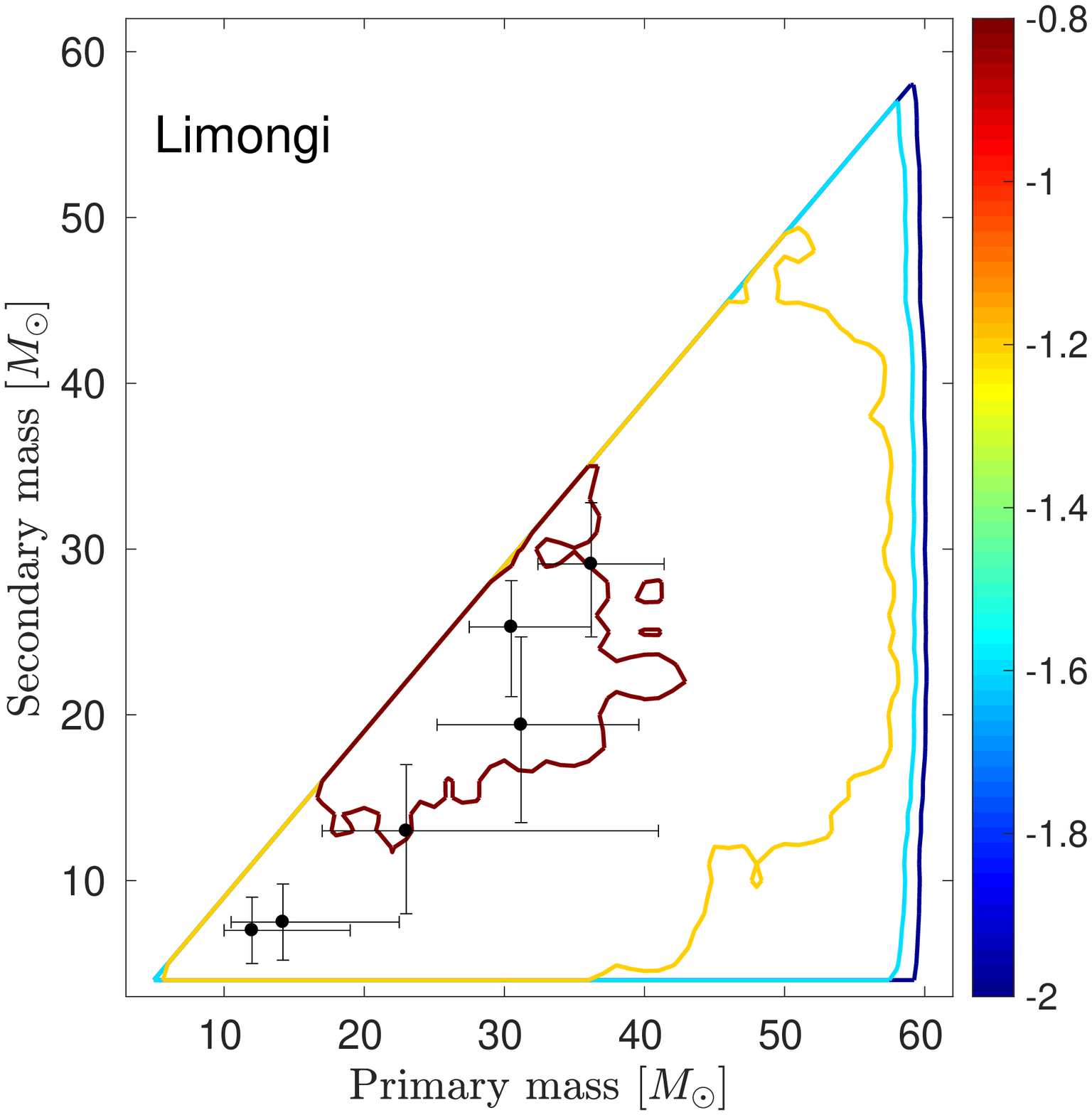, width=.3\textwidth}&
\epsfig{file=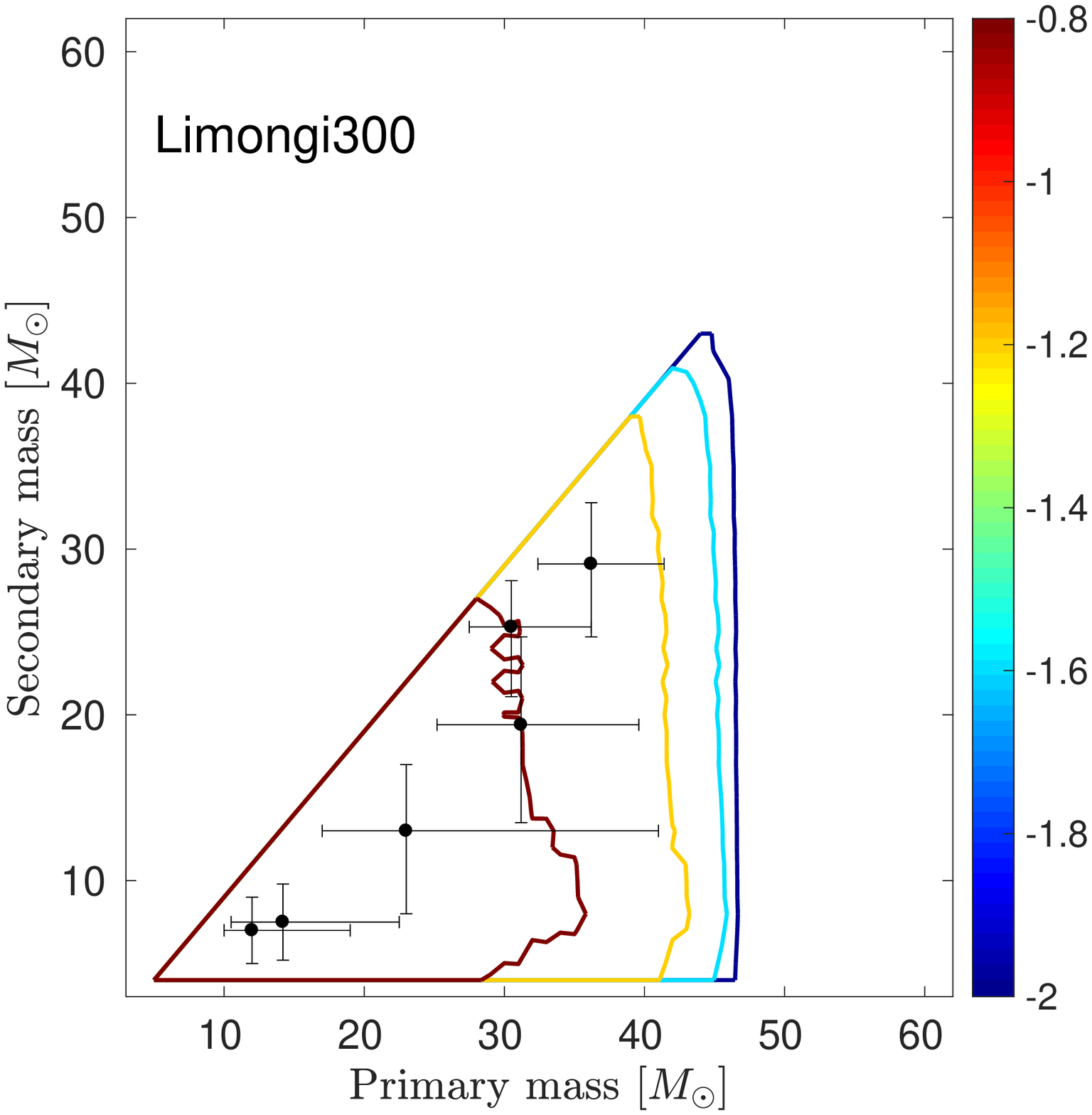, width=.3\textwidth}
\end{tabular}
\caption{Mass distribution of merging binary BHs in the astrophysical models considered in this paper: \emph{Fryer} (upper left), \emph{WWp} (upper right), \emph{Limongi} for non-rotating stars (lower left) and \emph{Limongi} with all stars rotating at $300$ km/s (lower right). Color coded is $\log_{10}R_{\rm det}$ where $R_{\rm det}$ is the detection rate in units of $[M_{\odot}^{-2}yr^{-1}]$ (see Eq.~(\ref{eq:Rdetdef}). The data points correspond to the published LIGO detections (including one event slightly below the discovery threshold). All the masses are in the source frame.}
\label{fig:MMdist}
\end{figure*}

\section{Detection rates of binary BH mergers}
\label{sec:res}

\subsection{Stellar-origin BHs}

The simplest way to compare between the four stellar evolution models discussed above is to calculate the detection rate of binary BH mergers that is implied from the detections made during the Advanced LIGO observing runs, as outlined in Section \ref{sec:rates_calc}. Specifically, we calculate the rate based on the O1 observing run. As can be seen in Table~\ref{tab:rates}, these range from $15$ to $59$ Gpc$^{-3}$yr$^{-1}$ and are in all the cases smaller than the one obtained in \citet{2016PhRvX...6d1015A} ($97^{+135}_{-67}$ Gpc$^{-3}$yr$^{-1}$ for their power-law model). Several factors could contribute to this discrepancy. First, \citet{2016PhRvX...6d1015A} assume that the sources are distributed uniformly in comoving volume, whereas our model predicts a specific redshift evolution that peaks at $z\sim 2$ \citep{2016MNRAS.461.3877D}. Therefore our model predicts lower relative numbers of low-redshift sources. Second, the BH mass function in our models differs from the one in \citet{2016PhRvX...6d1015A} because here we examine various initial mass-remnant mass relations, as discussed above. Finally, we note that our treatment of the selection function $f(\theta)$ is oversimplified with respect to the analysis of \citet{2016PhRvX...6d1015A}. In view of the uncertainty in the astrophysical model, it is also unclear which of these interpretations is correct, but it is important to keep in mind that the merger rates computed from the observed number of events are model-dependent. These results may be important for predicting the level of the expected stochastic background of GW \citep{2016PhRvL.116m1102A}, although we note that the observational uncertainties, due to the small number of events, are still more significant that the modeling uncertainty.

\begin{table}
\begin{center}
\caption{Merger rates deduced from LIGO O1 observations assuming different astrophysical models (see text for discussion).}

\begin{tabular}{c|c}

 & Rate $[Gpc^{-3}yr^{-1}]$ \\
 \hline
Fryer & 18 \\
WWp & 59 \\
Limongi & 15 \\
Limongi300 & 32 \\
\hline
\label{tab:rates}
\end{tabular}
\end{center}
\end{table}

In Figure~\ref{fig:MMdist} we plot the contours of constant detection rates per unit mass squared (in units of events per yr per $M_{\odot}^2$) for each of our models in the $M_1-M_2$ plane, where $M_1$ and $M_2$ are the primary and secondary BH masses, respectively (see Eq.~(\ref{eq:Rdetdef})). We also show the events detected by LIGO by black points with error bars. For example, comparing the first LIGO detection GW150914 with our models, we see that the \emph{Fryer} model predicts $\sim 0.16$ such detections per year per $M_{\odot}^2$ which, taking into account the error bars on the observed masses and the O1 coincident analysis time of $51.5$ days, gives $\sim 1$ expected detections in this model. The same calculation applies to the $Limongi$ models, but the $WWp$ case clearly produces too few BHs above $\sim 25M_{\odot}$. It is important to mention that these results depend on our model parameter $\beta$ (the number of BHs that are in binaries and that merge within a Hubble time). However we stress that the relative mass distribution is not affected by our choice of $\beta$ as long as it is taken to be a constant. Our value $\beta=0.01$ was chosen to roughly correspond to most of the models considered here. The only exception is $WWp$ which cannot be accommodated even with the maximal (and unrealistic) value of $\beta=1$. 

However, the most interesting (and robust) conclusion from our calculation is that the models discussed here present various specific features in their mass distributions of detectable BHs. For example, the \emph{WWp} and the \emph{Limongi300} models produce negligible number densities of BHs with masses above $\sim 25M_{\odot}$ and $\sim 45M_{\odot}$, respectively. This means that these models can be excluded even with a very small number of detections of 'heavy' sources. 

The case of the \emph{Fryer} and \emph{Limongi} models is even more striking: while they produce nearly identical \emph{total} numbers of detectable mergers, the mass distribution of these events is quite different. Specifically, in the \emph{Fryer} model the detectable binaries tend to cluster around $\sim 20-30M_{\odot}$. This feature of the \emph{Fryer} model can be traced back to the fact that in this description more massive stars experience stronger winds in such a way as to create an accumulation of BH masses at $\sim 20-30M_{\odot}$, as can be seen from Fig.~4 and eq. (7) in \citet{2012ApJ...749...91F} and Figure \ref{fig:models} above. On the other hand, the mass distribution of \emph{detectable} sources in the \emph{Limongi} model is predicted to be almost flat, with the exception of a small 'island' at $M\sim 20-40M_{\odot}$, possibly because in this model the reduced number densities of more massive BHs are roughly compensated by the fact that they are easier to detect. In this case, a large number of detections of $\sim 10M_{\odot}$ sources will probably exclude the \emph{Fryer} model while favouring the \emph{Limongi} model.

With only $5+1$ detected events, we clearly cannot rule out any of these models, but it may be possible when the number of detections increases. We can then study their mass distribution looking for specific features: do the sources cluster around specific mass values? Is there a mass cutoff? In particular, it might be interesting to estimate the number of detections necessary to rule out specific models, and we plan to do it in an upcoming paper. A possible caveat to this approach is that several channels for BH formation (i.e. primordial BHs, PopIII remnants, dynamical formation) may co-exist, rendering the distribution even more complex. 

We note that in our approach, the galaxy evolution processes, including the SFR and the metallicity evolution are the same for all the models, and the differences in the resulting distribution of BH masses can be directly attributed to differences in the stellar evolution model. On the other hand, our framework gives us the ability to marginalize over the unknown astrophysical parameters.

\begin{figure*}
\begin{tabular}{cc}
\epsfig{file=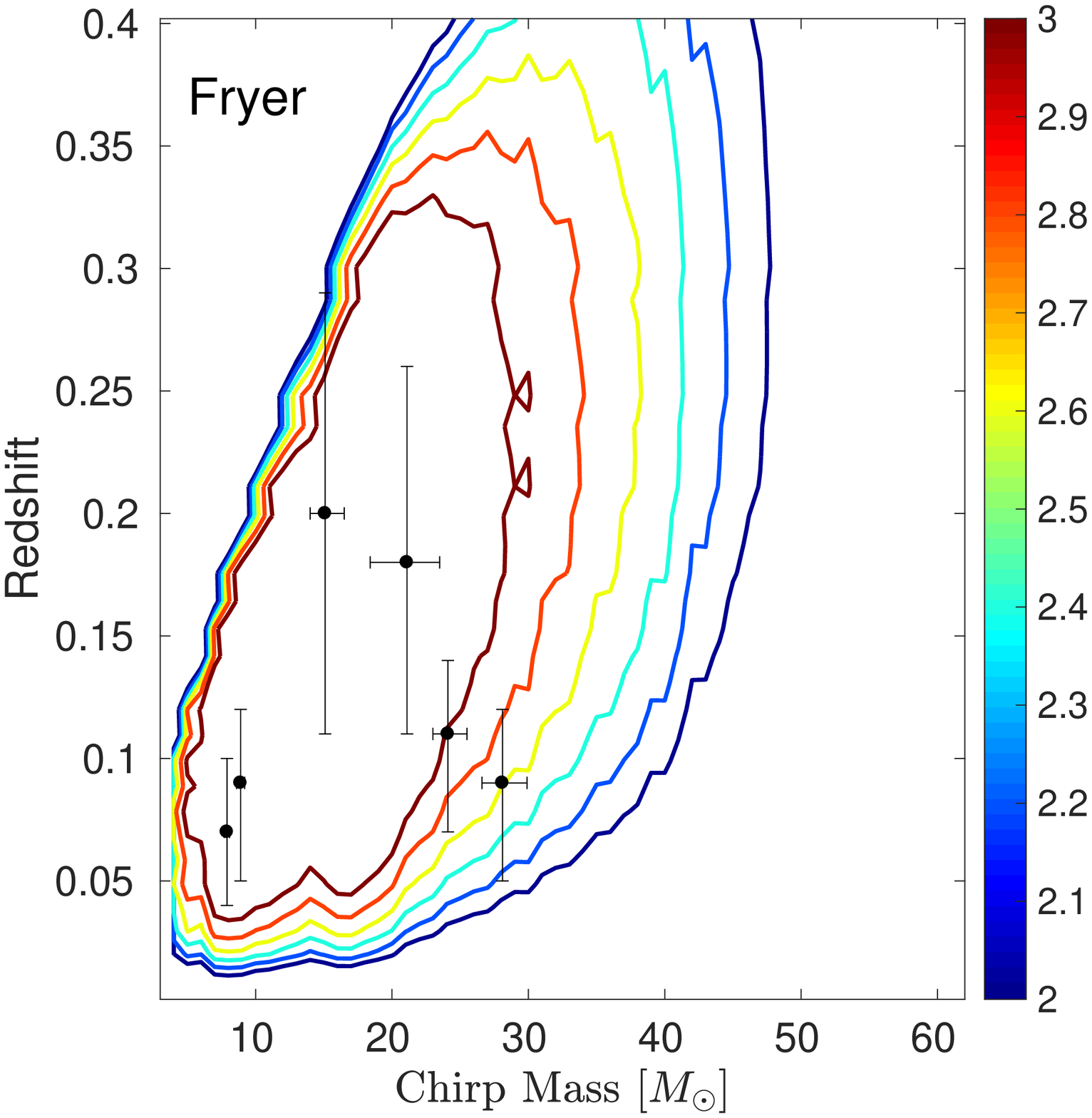, width=.3\textwidth}&
\epsfig{file=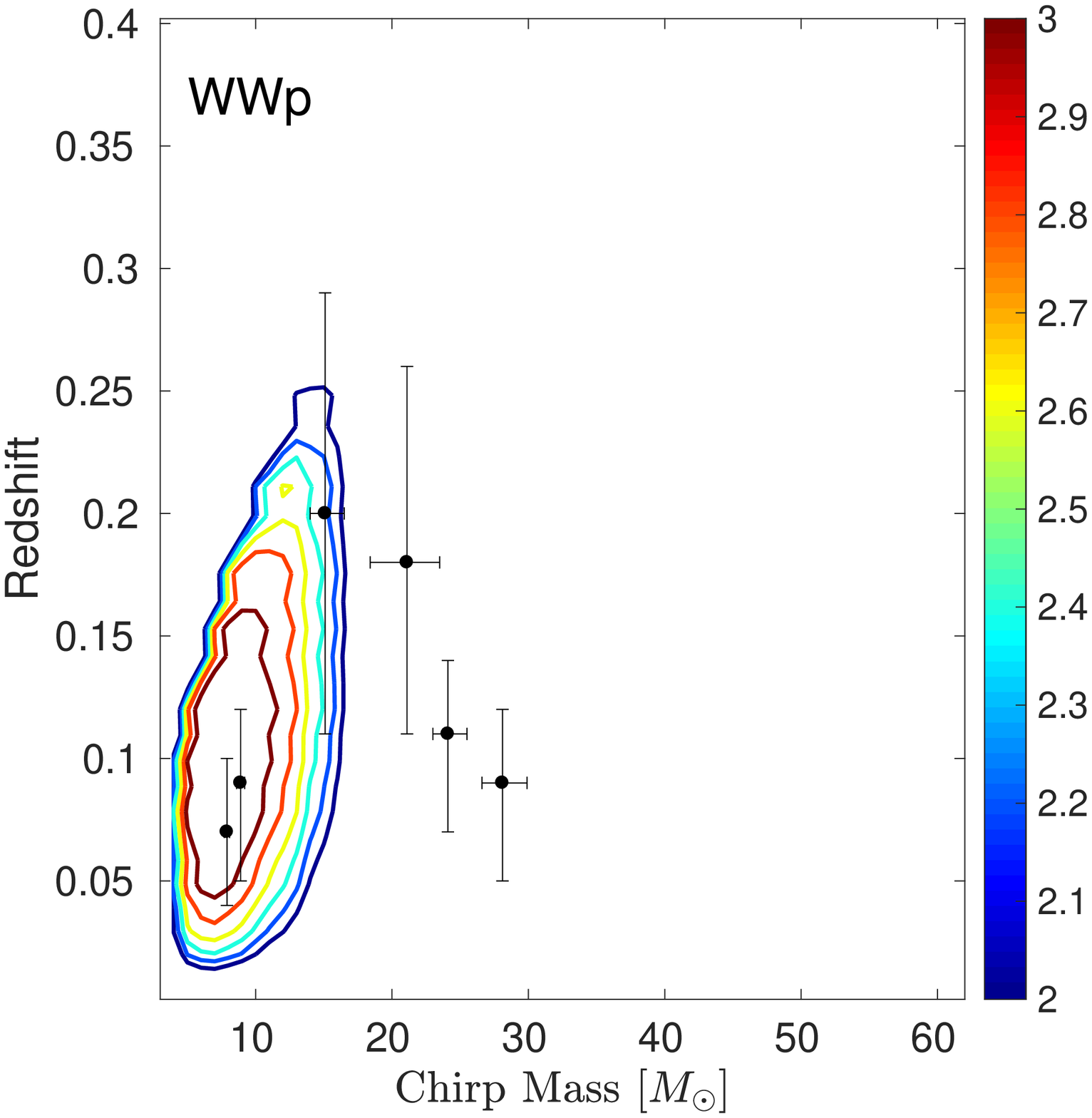, width=.3\textwidth}\\
\epsfig{file=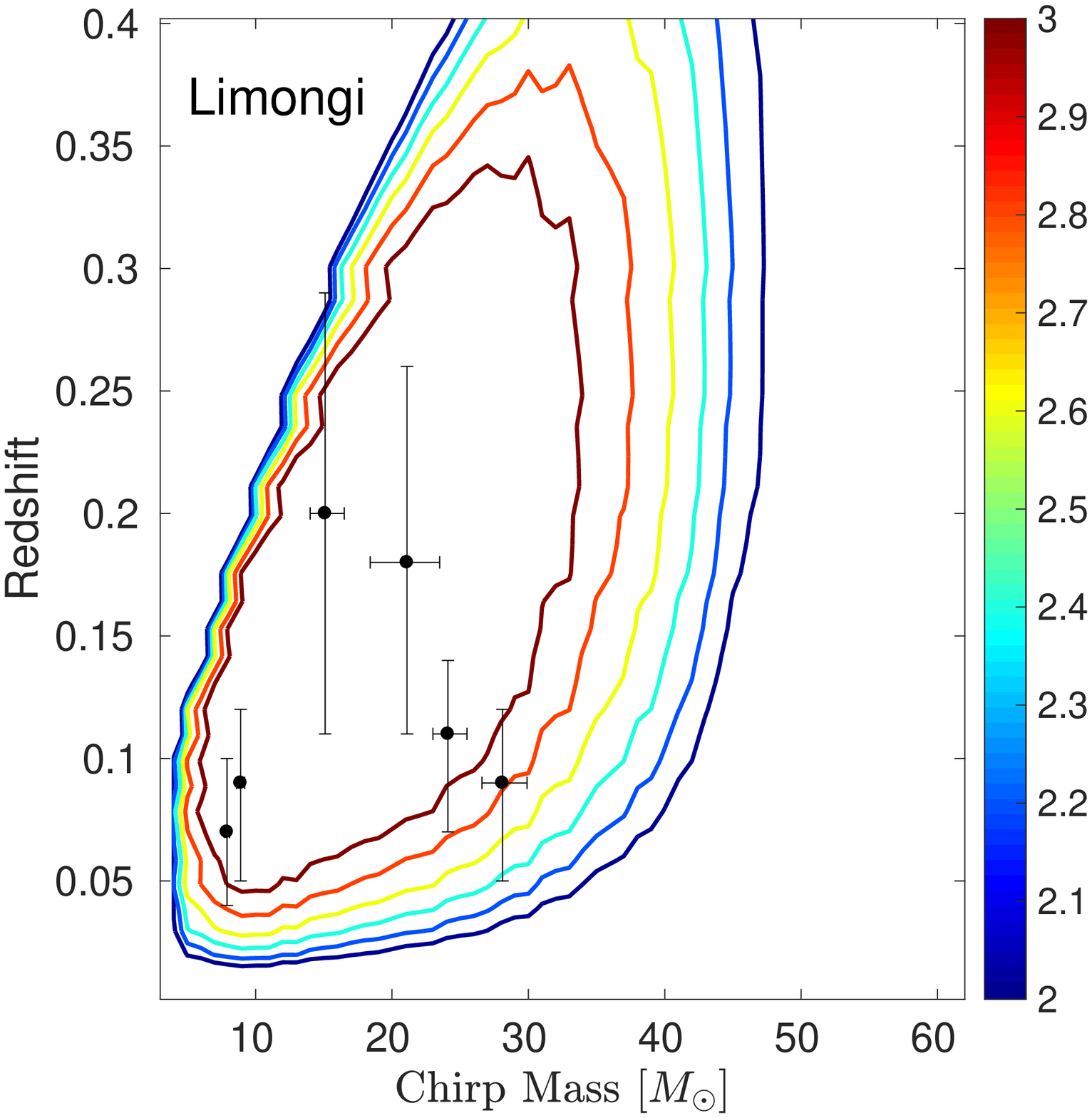, width=.3\textwidth}&
\epsfig{file=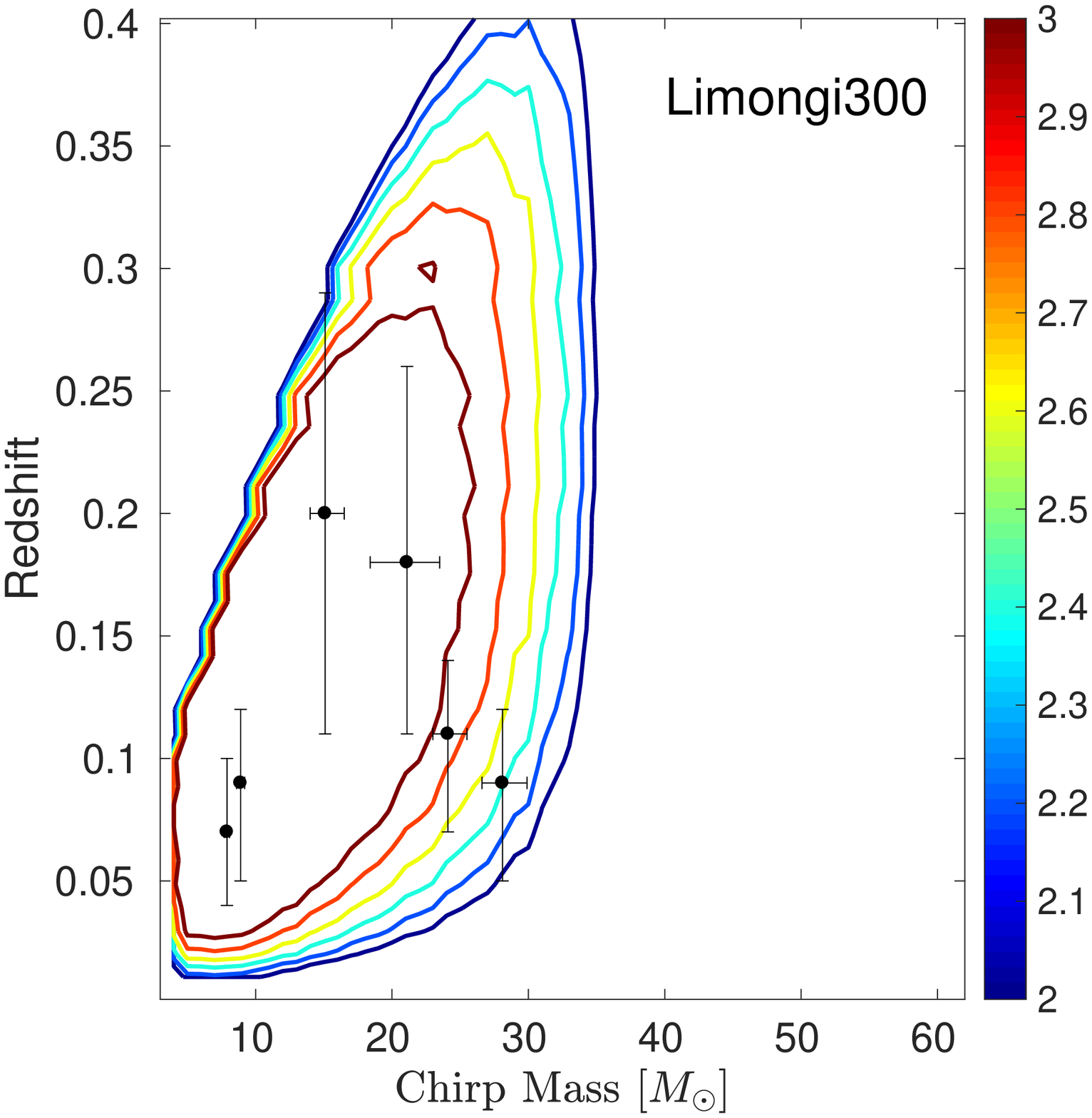, width=.3\textwidth}
\end{tabular}
\caption{Distribution of detectable events in the merger redshift-chirp mass plane for the astrophysical models considered in this work. Color coded is $\log_{10}(R'_{\rm det})$ in units of events per unit solar mass per unit redshift. The data points correspond to the published LIGO detections (including one event slightly below the discovery threshold). All the masses are in the source frame.}
\label{fig:MZdist}
\end{figure*}

In addition to the distribution of detectable sources in the $M_1-M_2$ plane we can calculate their redshifts. Fig. \ref{fig:MZdist} shows the contours of constant detection rates $R'_{\rm det}$ in the $M_c-z$ plane, where the chirp mass is $M_c=(M_1M_2)^{3/5}/(M_1+M_2)^{1/5}$ and
\begin{multline}
 R'_{\rm det}(M_c,z)  = \\
 \int R_{\rm det}(M_1,M_2,t)\delta\left(M_c(M_1,M_2)-M_c\right) \dd M_1 \dd M_2\left|\frac{\dd t}{\dd z}\right|\:.
\end{multline} 
Combining these predictions with those from Fig.~\ref{fig:MMdist} will result in even tighter constraints. As in the case of the $M_1-M_2$ plane, the \emph{Fryer} and \emph{Limongi} models seems to provide a better correspondence to LIGO detections.

\begin{figure}
\epsfig{file=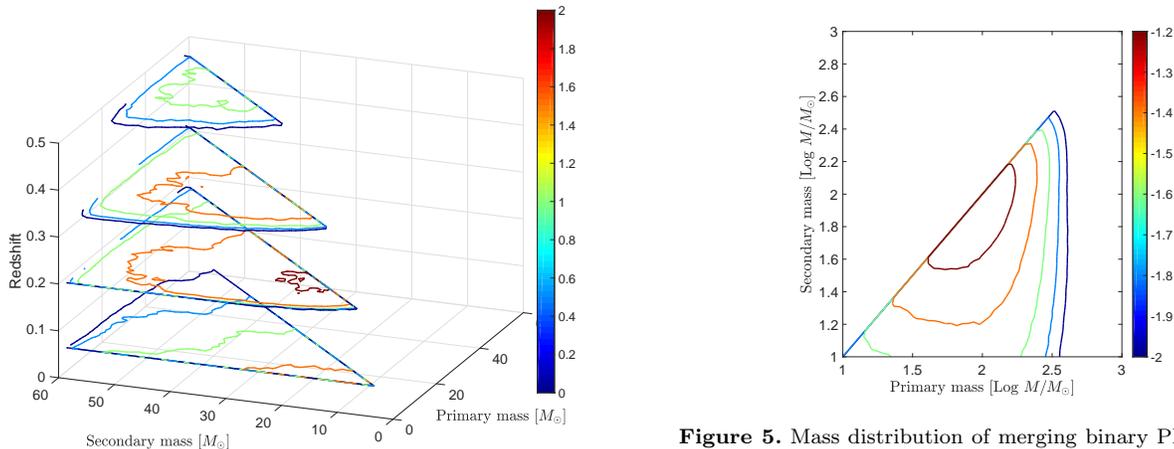, width=.45\textwidth}
\caption{Distribution of detectable events in the merger redshift-primary mass-secondary mass plane for the \emph{Limongi} model. Color coded is the detection rate in units of $[M_{\odot}^{-2}yr^{-1}]$ per unit of redshift. All the masses are in the source frame.}
\label{fig:MMZdist}
\end{figure}

Finally, we show the full 3D distribution in the primary mass - secondary mass - redshift plane on Figure \ref{fig:MMZdist} for the \emph{Limongi} model. As can be seen, most of the contribution comes from $z\sim 0.2-0.3$. 

\subsection{Primordial BHs}
\begin{figure}
\centering
 \epsfig{file=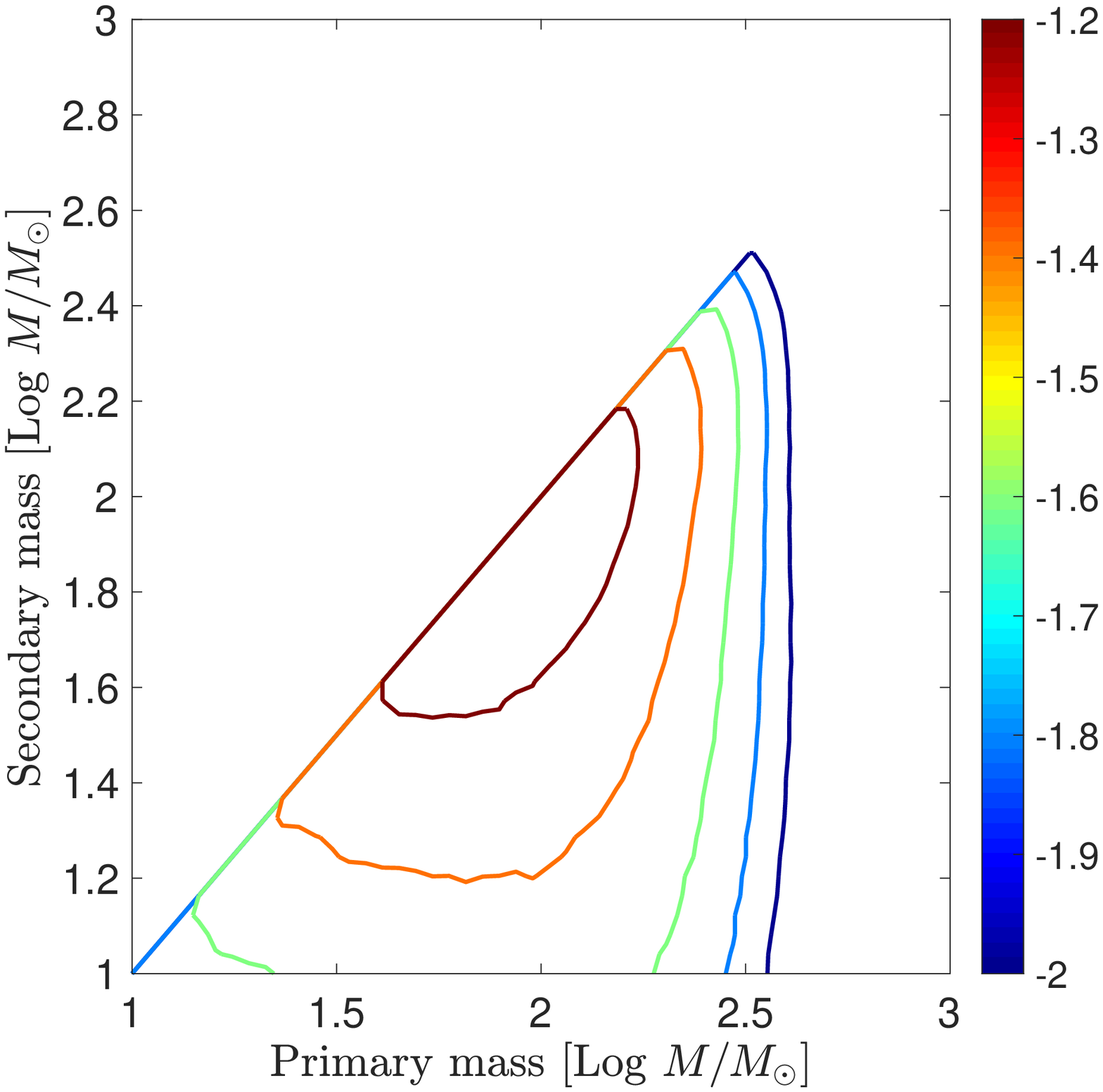,width=0.3\textwidth}
 \caption{Mass distribution of merging binary PBHs using the phenomenological model discussed in the text, namely $10\%$ of dark matter in PBHs, $P(t)\propto 1/t$ merging time delay distribution, efficiency of binary formation of $10^{-2}$ per Gyr for $z<2$ and a power-law mass function with slope $\alpha=0.5$ in the mass range $(10-1000)M_{\odot}$. Color coded is $\log_{10}R_{\rm det}$ where $R_{\rm det}$ is the detection rate in units of $[M_{\odot}^{-2}yr^{-1}]$.}
 \label{fig:pbh}
\end{figure}

We consider a generic PBH model with $\Gamma_{\rm PBH}=10^{-2}$~Gyr$^{-1}$ for $z<2$ in eq. (\ref{eq:Gpbh}), mass range $M_{\rm min}=10M_{\odot}$, $M_{\rm max}=1000M_{\odot}$, fraction of PBHs as dark matter $\Omega_{\rm PBH}/\Omega_{\rm DM}=q=0.01$ and slope of their mass function $\alpha=0.5$. These values were chosen to demonstrate the potential differences of this population from stellar-origin BHs, in particular, we chose a very shallow mass distribution to give more weight for high-mass BHs that can form in this scenario. The detection rates that would be obtained by LIGO in this case are shown in Figure \ref{fig:pbh}. The shallow slope of the PBH mass distribution results in a relatively high detection rate of BHs in the PISN mass gap $M\gtrsim 60M_{\odot}$, in particular a peak around $\sim 100M_{\odot}$. It seems tempting to suggest that even a single detection of such BH masses would provide a strong hint towards a primordial origin, although more detailed studies are needed in order to exclude other formation scenarios such as the dynamical formation channel. Further work is needed to relate the phenomenological model described here to detailed PBH binary formation scenarios.

\section{Parameter estimation}
\label{sec:stats}

In this Section we demonstrate the potential power of our approach by estimating some of the model parameters with maximal likelihood analysis. The number of detections required to estimate model parameters was discussed in previous studies \citep[e.g.][]{2016A&A...594A..97B,2017arXiv170901943W,2017arXiv171106287B} and it was shown that in general between a few hundreds to a thousand detections will be needed. The main difficulty in treating this issue is the choice of parameters which vary among different models.  In the approach developed in this paper some of parameters are common among the models, which facilitates model comparison. 
In the following we focus on two parameters: $\beta$, the fraction of BHs that are in binaries that merge within a Hubble time, and $\gamma$, the power-law of the merger delay time distribution (see Eq. (\ref{eq:tdist})). It is useful to vary also other model parameters, especially the ill-constrained shape of the mass distribution (Eq. (\ref{eq:Pmm})), and possibly the parameters of the galaxy evolution model, such as the SFR and the IMF. In view of the large number of parameters, a full analysis necessitates a Monte Carlo Markov chain approach, which we leave to a follow-up study. 

The accuracy of our analysis depends crucially on the measurement precision. Here we choose to focus on the chirp mass, which was measured to very good precision in O1 and O2 LIGO/Virgo runs. In the future other observables can be included, such as the individual BH masses and redshifts. 

\begin{figure}
\begin{tabular}{c}
\epsfig{file=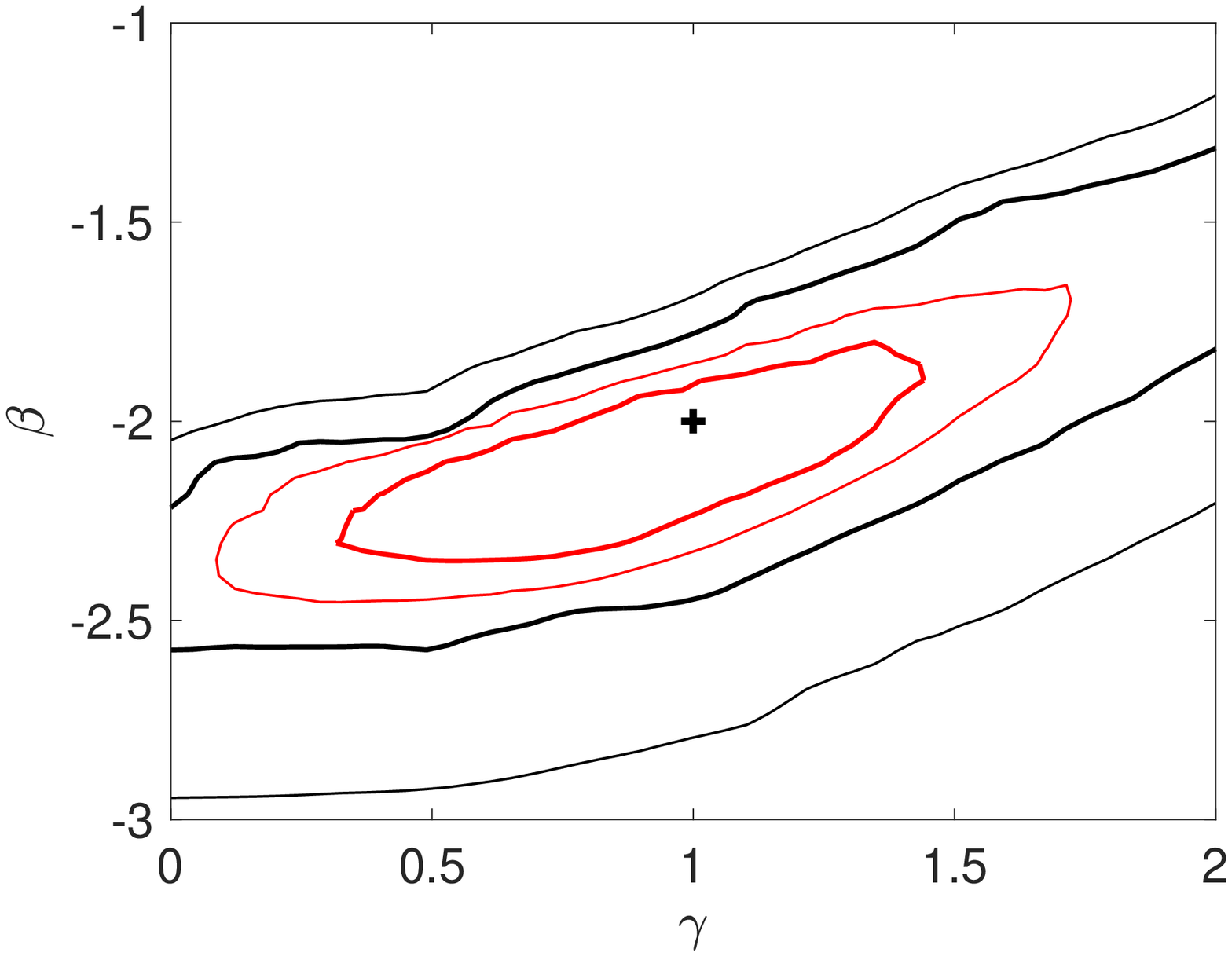, width=.4\textwidth}\\
\epsfig{file=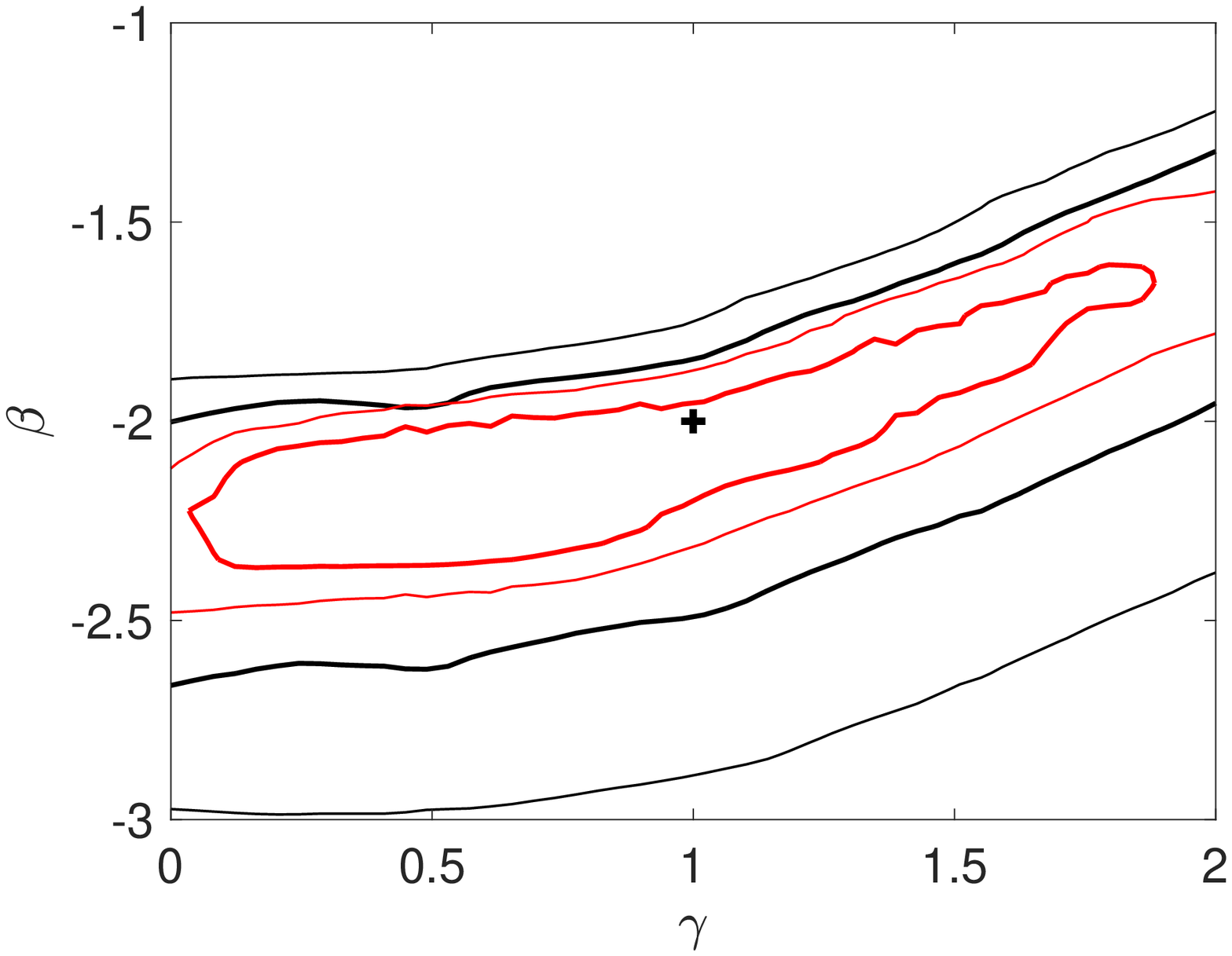, width=.4\textwidth}
\end{tabular}
\caption{Constraints on model parameters $\beta$, the fraction of BHs that are in binaries that merge within a Hubble time, and $\gamma$, the power-law of the merger delay time distribution with $100$ and $500$ events (black and red curves, respectively), using only measurements of the chirp mass for the \emph{Fryer} model (upper panel) and the \emph{Limongi} model (lower panel). Shown are the $2\sigma$ and $5\sigma$ contours. Note the degeneracy between the two parametres.}
\label{fig:mparams}
\end{figure}

When testing a given model we calculate the detection rate per unit time and per unit chirp mass assuming some fiducial values of $\beta$ and $\gamma$, as outlined above, which gives us the mean expected number of detections $\dd N_{\rm det}/\dd M_c$ made during a given observation time $T_{\rm obs}$. We then choose $T_{\rm obs}$ that corresponds to $N_{\rm tot}=100$ and $N_{\rm tot}=500$ total detectable events (black and red curves on Fig. \ref{fig:mparams}, respectively). We bin our results in mass bins of width $3M_{\odot}$ to obtain $\Delta N_{\rm det}$ and ignore the errors on the masses (that is, we assume that the errors are much smaller than the width of each bin, and ignore the cross-correlations between the bins). For each bin we draw a number from a Poisson distribution with mean $\Delta N_{\rm det}$ to obtain the \emph{mock observations}. We then compare these mock observations to the mean expected number of detections for the model and the set of parameters we wish to test. In particular, we assume flat priors on $\gamma$ and Log$(\beta)$ and perform a maximum likelihood analysis.

To generate our mock observations we use the fiducial values $\beta=0.01$ and $\gamma=1$ and either the \emph{Fryer} or \emph{Limongi} model. As can be seen in Fig. \ref{fig:mparams}, $\beta$ and $\gamma$ are highly degenerate. This is to be expected: in the case of a steep merger time delay distribution (large $\gamma$) most of the BH binaries merge immediately after formation, close to the peak of star formation around $z\sim 2$ and are thus not observed. To reach the same overall number of detected mergers we need to have a larger fraction $\beta$ of binaries that are in binaries and are on close enough orbits to merge within a Hubble time. 


\section{Discussion}
\label{sec:dis}

The discovery of GW from merging binary BHs opens new perspectives for the studies of stellar evolution and BH formation. In this paper we introduced a framework that can be used to analyze upcoming GW detections in a full astrophysical context with the aim of constraining stellar evolution models. We qualitatively showed the effect of different models on the mass and redshift distribution of potential LIGO sources. We find that among the stellar evolution models discussed here the \emph{Limongi} model without stellar rotation and the \emph{Fryer} model provide the best description of the observed distribution. These models differ in the mass distribution of \emph{detectable} BHs: while the \emph{Fryer} model predicts a concentration of BHs around $\sim 20-30M_{\odot}$ (a result of the modeling of mass loss in this case), the distribution is almost flat in the \emph{Limongi} model. It therefore seems possible to discriminate between these models with more observations of BH mergers. We also find that the \emph{WWp} model is not compatible with LIGO detections since it produces too few BHs above $\sim 25M_{\odot}$. Moreover, the \emph{Limongi300} model, in which all the stars rotate at $300$ km/sec is also unlikely due to a cutoff it introduces at $\sim 45M_{\odot}$, which is a result of the fact that rotating stars undergo PISN at lower masses than their non-rotating counterparts.


We also performed a basic parameter estimation analysis, focusing only on $\beta$ and $\gamma$  and using $100$ events drawn from a population computed with either \emph{Fryer} or \emph{Limongi} models. We found that these parameters are degenerate: the same number of detections can be obtained for \emph{lower} binary fraction and \emph{shallower} time delay distribution. 

It will be interesting to consider alternative BH formation channels, such as the dynamical formation channel, PopIII remnants and primordial BHs. In view of our results, it is clear that models which present specific unique features in their mass and/or redshift distribution will be the easiest to constrain. For example, even a single $\sim 150M_{\odot}$ BH could point to one of these alternative channels, since it cannot be produced via standard stellar evolution (as it would fall in the PISN range). However, in the absence of such 'smoking-gun' detections and in view of the large variety of stellar evolution models it might be difficult to constrain some of these alternative channels with current ground-based interferometers. For example, the generic primordial BH scenario, discussed in this paper, seems to be difficult to constrain if the merger times are distributed roughly like $1/t_{\rm delay}$ as in \citet{2016PhRvL.117f1101S} (and similarly to the stellar-origin BHs), and the BH mass function is bottom-heavy with a cutoff at $\sim 70M_{\odot}$ as in \citet{2017arXiv170603746C}. While the redshift distribution of these sources will be constant out to high redshifts, contrary to the case of stellar-origin BHs, this feature will not be detectable before the next generation of ground-based interferometers becomes operational.

Finally, we have not discussed the spins of the merging BHs, which can provide additional constraints, in particular for the dynamical formation channel, and which we plan to include in future work. 

\section*{Acknowledgments}                                                      
We thank the anonymous referee for useful suggestions that helped improve the manuscript. ID is grateful to Thibaut Louis for useful discussions. This work has been done within the Labex ILP (reference ANR-10-LABX-63), part of the Idex SUPER, and received financial state aid managed by the Agence Nationale de la Recherche, as part of the programme Investissements d'avenir under the reference ANR-11-IDEX-0004-02. We acknowledge the financial support from the EMERGENCE 2016 project, Sorbonne Universit\'{e}s, convention no. SU-16-R-EMR-61 (MODOG).

\bibliographystyle{mn2e}
\bibliography{detrates}

%

\end{document}